\documentclass[pra,twocolumn,showpacs,superscriptaddress]{revtex4}%
\usepackage[dvips]{graphics, color}
\usepackage{amssymb}
\usepackage{amsmath}
\usepackage{graphicx}
\usepackage{epsfig}
\usepackage{amsfonts,times}
\usepackage{amsfonts}%
\begin{document}
\title{Theory of Optically-Driven Sideband Cooling for Atomic Collective
Excitations and Its Generalization}
\author{Yong Li}
\affiliation{Department of Physics and Center of Theoretical and Computational Physics, The
University of Hong Kong, Pokfulam Road, Hong Kong, China}
\author{Z. D. Wang}
\affiliation{Department of Physics and Center of Theoretical and Computational Physics, The
University of Hong Kong, Pokfulam Road, Hong Kong, China}
\author{C. P. Sun}
\affiliation{Institute of Theoretical Physics, The Chinese Academy of Sciences, Beijing,
100190, China}
\date{\today }

\begin{abstract}
We explore how to cool atomic collective excitations in an
optically-driven three-level atomic ensemble, which may be described
by a model of coupled two harmonic oscillators (HOs) with a
time-dependent coupling. Moreover, the coupled two-HO model is
further generalized to address other cooling issues, where the
lower-frequency HO can be cooled whenever the cooling process
dominates over the heating one during the sideband transitions.
Unusually, due to the absence of the heating process, the optimal
cooling of our first cooling protocol for collective excitations in
an atomic ensemble could break a usual sideband cooling limit for
general coupled two-HO models.

\end{abstract}

\pacs{03.65.-w,
37.10.De,  
43.58.Wc  
} \maketitle

\emph{Introduction.- } Recently, quantum information processing
based on collective excitations in atomic ensembles has attracted
more and more attentions. Photons are good carriers of quantum
information due to their fast velocity and low leakage, while may
not be easy to store. Naturally, it is desired to study atomic
ensembles as potential quantum memory units of photons due to the
long coherence time. Interestingly, the form-stable dark-state
polariton (DSP) \cite{Fleischhauer2000} associated with the
propagation of quantum optical fields via electromagnetically
induced transparency (EIT) \cite{Harris-EIT}, was proposed in a
three-level $\Lambda$-type atomic ensemble. In the low excitations
limit, DSP can be described as a hybrid bosonic mode \cite{Sun2003}.
By controlling the mixing angle between light and matter components
of DSP, the optical pulse can be decelerated and \textquotedblleft
trapped\textquotedblright\ via mapping its shape and quantum state
onto meta-stable collective-excitation state of matter. That means
the quantum information storage
\cite{Fleischhauer2000,Sun2003,Fleischhauer and Lukin} can be
achieved in the atomic ensembles by adiabatically controlling the
coupling.

As is known, collective excitations could also be used in quantum
communication in atomic ensembles and linear optics. Since the work of
Duan-Lukin-Cirac-Zoller~\cite{DLCZ},
a number of protocols \cite{repeater,storage,Liu,ZhaoBo2007} have
been proposed to implement robust long-distance quantum
communications, quantum repeaters, and quantum information storages
based on atomic ensembles over long photonic lossy channels.

In a realistic atomic ensemble, a given collective-excitation mode may have a
finite thermal population due to the interaction with the thermal bath at
finite temperature. This means that it is necessary to cool the thermal
excitations in quantum information processing based on atomic collective
excitations. In this Letter, we consider a driven three-level atomic ensemble
that can be modeled by coupled two harmonic resonators (HOs), and then
elaborate how to cool the low-frequency collective-excitation mode near its
ground state in this kind of systems.

On the other hand, various nano- (or submicron-) mechanical
resonators have been investigated~\cite{nano} extensively\textbf{
}in recent years. To reveal the quantum effect in the
nano-mechanical devices, various cooling schemes
\cite{MR-cooling,radiation-pressure cooling
experiment,Wilson-Rae:2007,Marquardt:2007,Genes:2008,Rabl2009} were
proposed to drive them to reach the standard quantum limit (SQL)
\cite{SQL}. A famous one among them is the optical
radiation-pressure cooling scheme \cite{radiation-pressure cooling
experiment} attributed to the (resolved) sideband cooling
\cite{Wilson-Rae:2007,Marquardt:2007,Genes:2008,Rabl2009}, which was
previously well-developed to cool the spatial motion of the trapped
ions \cite{trap-ion-cooling} or the neutral atoms
\cite{atom-cooling}. Notably, our cooling scheme for atomic
ensembles is based on the sideband structure induced by the
lower-frequency mode, which is time-dependently coupled with the
higher-frequency mode to loss its energy. Moreover, we generalize
the above coupled two-HO model to other two types of cooling model
beyond the optical radiation-pressure cooling of mechanical
resonator. In the generalized model, the lower-frequency HO can be
cooled with a usual sideband cooling limit, whose cooling mechanism
can also be employed to understand the cooling of collective
excitations in the atomic ensembles. It is remarkable that our
protocol of atomic ensemble breaks the limit of usual sideband
cooling due to the absence of counter-rotating terms in such a
coupled two-HO model.

\begin{figure}[th]
\includegraphics[width=8cm]{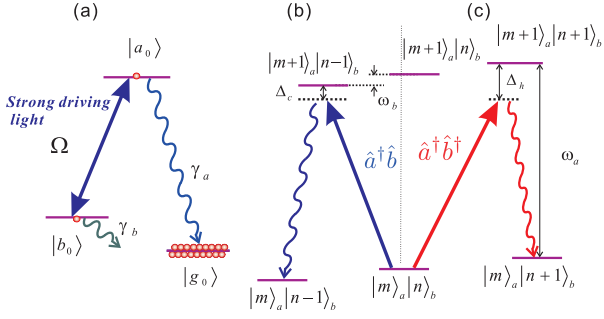}\caption{(Color
online) (a) Three-level atomic ensemble with most atoms staying in
the ground states $\left\vert g_{0}\right\rangle $. The strong
driving light couples to the transition from the meta-stable state
$\left\vert b_{0}\right\rangle $ to the excited one $\left\vert
a_{0}\right\rangle $ for each atom. The electric-dipole transition
$\left\vert g_{0}\right\rangle $$\rightarrow$$\left\vert
a_{0}\right\rangle $ is permitted, but $\left\vert
g_{0}\right\rangle $$\rightarrow$$\left\vert b_{0}\right\rangle $ is
forbidden. The waved lines denote the decay processes with
$\gamma_{a,b}$ the corresponding decay rates. (b) The cooling
process ($\left\vert n\right\rangle _{b}\rightarrow\left\vert
n-1\right\rangle _{b}$) and (c) the heating process ($\left\vert
n\right\rangle _{b}\rightarrow \left\vert n+1\right\rangle _{b}$)
for mode $b$ starting form $\left\vert m\right\rangle _{a}\left\vert
n\right\rangle _{b}$ in the sideband structure forming by splitting
a-mode with the low-frequency b-mode. $\Delta_{c}$
($\equiv\omega_{a}-\omega_{b}-\omega_{d})$ and $\Delta_{h}$
($\equiv\omega _{a}+\omega_{b}-\omega_{d})$ are the detunings for
the anti-Stokes (cooling) and Stokes (heating) transitions,
respectively. } \label{atomic-ensemble}
\end{figure}

\emph{Three-Level atomic ensemble modeled by two coupled
oscillators.- } Let us consider an ensemble of $N$ identical
three-level atoms as seen in Fig.~\ref{atomic-ensemble}(a). A strong
classical driving light field is homogenously coupled to each atomic
transition from the metastable state $\left\vert b_{0}\right\rangle
$ to the excited one $\left\vert a_{0} \right\rangle $. Then the
Hamiltonian reads ($\hbar=1$ hereafter)
\begin{equation}
H=\omega_{a}\sum_{i=1}^{N}\sigma_{a_{0}a_{0}}^{(i)}+\omega_{b}\sum_{i=1}^{N}\sigma
_{b_{0}b_{0}}^{(i)}+(\Omega
e^{i\omega_{d}t}\sum_{i=1}^{N}\sigma_{b_{0}a_{0}} ^{(i)}+h.c.),
\label{H-1}
\end{equation}
where $\omega_{g,a,b}$ are the corresponding energies of the atomic states
$\left\vert g_{0}\right\rangle $, $\left\vert a_{0}\right\rangle $ and
$\left\vert b_{0}\right\rangle $ respectively, and the ground state energy
$\omega_{g}=0$. $\Omega$ is the coupling strength of the driving light field
(with the carrier frequency $\omega_{d}$), which can be assumed to be real.
%

Normally, a weak quantized probe light would couple to the transition
$\left\vert g_{0}\right\rangle $$\rightarrow$$\left\vert a_{0}\right\rangle $.
Thus, a so-called $\Lambda$-type three-level atomic ensemble configuration can
be constructed associated with the well-known EIT and group-velocity slowdown
phenomena. In such an ensemble, the DSP can also be obtained as the
superposition of the optical mode and the atomic collective-excitation mode
\cite{Fleischhauer2000,Sun2003,Fleischhauer and Lukin}. Based on the notations
of EIT and DSP, the atomic ensemble can be a unit of quantum memory and be
used to store the quantum information of, e.g., the photons. Here, we focus
only on the cooling of the atomic collective excitations in the absence of the
probe light field, also noting that extensive studies have been made in the
framework of optically-pumping an individual atom into its internal
lowest-energy ground state~\cite{pump}.

We now introduce the bosonic operators
$\hat{a}=\sum_{i}\hat{\sigma}_{g_{0}a_{0}}^{(i)} /\sqrt{N}$ and
$\hat{b}=\sum_{i}\hat{\sigma}_{g_{0}b_{0}}^{(i)}/\sqrt{N}$ for
atomic collective excitations~\cite{Sun2003,sun-excitation}, which
satisfy $[\hat{a},\hat{a}^{\dagger}]=1$,
$[\hat{b},\hat{b}^{\dagger}]=1$ and
$[\hat{a},\hat{b}^{\dagger}]=0=[\hat {a},\hat{b}]$
in the limit of $N\rightarrow\infty$ with low excitations. Then,
Hamiltonian (\ref{H-1}) is modeled by the coupled two-HO model, and
can be further rewritten in a time-independent form in the rotating
framework as
\begin{equation}
H_{I}=\Delta\hat{a}^{\dagger}\hat{a}+\omega_{b}\hat{b}^{\dagger}\hat{b}%
+\Omega(\hat{a}^{\dagger}\hat{b}+h.c.)\label{H-4}%
\end{equation}
with the detuning $\Delta\equiv\omega_{a}-\omega_{d}$. In the
derivation of the above Hamiltonian, we have used the rotating wave
approximation (RWA) when $\{\left\vert
\omega_{ab}-\omega_{d}\right\vert ,\left\vert \Omega\right\vert
\}\ll(\omega_{ab}+\omega_{d})$ (where
$\omega_{ab}\equiv\omega_{a}-\omega_{b} $), which is always
fulfilled for most realistic atoms.

\emph{Sideband cooling for atomic collective excitations.- }
Generally, the atomic collective-excitation modes have non-vanishing
mean thermal populations due to their couplings to the bath at
finite temperature. In experiments, the frequency of the
higher-frequency atomic collective-excitation, i.e., mode $a$, is of
the order of $2\pi\times10^{14}$ Hz, which implies that its mean
thermal excitation number can be considered as zero even at room
temperature. Usually, the atomic ground state $\left\vert
g_{0}\right\rangle $ and meta-stable one $\left\vert
b_{0}\right\rangle $ are selected as the atomic two hyperfine levels
with the frequency difference being the order of $2\pi\times10^{9}$
Hz. Although there is no optical dipole transition between
$\left\vert b_{0}\right\rangle $ and $\left\vert g_{0}\right\rangle
$ because of the electric dipole transition rule, the decay from
$\left\vert b_{0}\right\rangle $ to $\left\vert g_{0}\right\rangle $
still exists due to the atomic collision or some other cases, with a
very low decay rate. Such a very-low decay rate means that the
lower-energy mode $b$ possesses a long coherence time, which is just
a distinct advantage of using the atomic collective excitations as
quantum memory units. However, in consideration of the high initial
mean thermal population $\bar{n}_{b}
=[\exp(\omega_{b}/k_{B}T)-1]^{-1}\sim10^{4}\gg 1$ at room
temperature $T\sim300$ K (with $k_{B}$ the Boltzmann constant), it
is necessary to cool the atomic collective-excitation modes to their
ground states before quantum information processing based on atomic
collective excitations.

In the presence of noises, we may have the following Langevin
equation from Hamiltonian~(\ref{H-4})
\begin{equation}
\dot{\hat{C}}=-\Gamma_{C}\hat{C}+i\Omega\hat{C}^{\prime}+\hat{F}
_{C}(t),\label{Lan-equ}
\end{equation}
where $C,C^{\prime}=a,b$ ($C\neq C^{\prime}$),
$\Gamma_{a}=\gamma_{a} /{2}+i\Delta$ and
$\Gamma_{b}$$=\gamma_{b}/{2}+i\omega_{b}$. The noise operators are
described by the correlations $\langle\hat{F}_{C}^{\dagger
}(t)\hat{F}_{C}(t^{\prime})\rangle
=\gamma_{C}\bar{n}_{C}\delta(t-t^{\prime})$. Here, $\gamma_{a,b}$
are the decay rates of collective-excitation modes $a$ and $b$,
respectively (for simplicity, we adopt the same symbols as those of
the atomic levels $\left\vert a_{0}\right\rangle $ and $\left\vert
b_{0}\right\rangle $), and $\bar{n}_{a,b}=[\exp(\omega_{a,b}
/k_{B}T)-1]^{-1}$ are the corresponding initial thermal populations
with $T$ the initial temperature of the thermal bath. Although the
above quantum Langevin equation has vanishing steady state solutions
$\langle\hat{a}\rangle=\langle\hat{b}\rangle=0$, the corresponding
quantum rate equations for the excitation numbers $\hat{n}_{C}=\hat
{C}^{\dagger}\hat{C}$ ($C=a,b$) read
\begin{align}
\frac{d}{dt}\left\langle \hat{n}_{C}\right\rangle  &  =\gamma_{C}(\bar{n}%
_{C}-\left\langle \hat{n}_{C}\right\rangle )-\left(  i\Omega\langle\hat
{\Sigma}\rangle+h.c.\right)  ,\\
\frac{d}{dt}\langle\hat{\Sigma}\rangle &  =-\zeta\langle\hat{\Sigma}%
\rangle+ig(\left\langle \hat{n}_{b}\right\rangle -\left\langle \hat{n}%
_{a}\right\rangle ),
\end{align}
where $\hat{\Sigma}=\hat{a}^{\dagger}\hat{b}$ and $\zeta=(\gamma_{a}%
+\gamma_{b})/2+i(\omega_{b}-\Delta)$.
Here we have used the non-vanishing noise-based relations \cite{Scully-book}
$\langle F_{C}^{\dagger}(t)\hat{C}(t)\rangle=\gamma_{C}\bar{n}_{C}/2$.

The steady state solutions of the quantum rate equations give the variation of
final mean population $\bar{n}_{b}^{\mathrm{f}}=\langle(\hat{b}^{\dagger
}-\langle\hat{b}^{\dagger}\rangle)(\hat{b}-\langle\hat{b}\rangle
)\rangle_{\mathrm{ss}}\equiv\bar{n}_{b}-\xi(\bar{n}_{b}-\bar{n}_{a})$ with
\begin{equation}
\xi=\frac{\Omega^{2}\gamma_{a}(\gamma_{a}+\gamma_{b})}{(\gamma_{a}+\gamma
_{b})^{2}\left(  \Omega^{2}+\frac{\gamma_{a}\gamma_{b}}{4}\right)  +\gamma
_{a}\gamma_{b}(\Delta-\omega_{b})^{2}}.\nonumber
\end{equation}
Then, from the Bose-Einstein distribution, the effective temperature
$T_{\mathrm{eff}}$ of mode $b$ is expressed as
\begin{equation}
T_{\mathrm{eff}}=\frac{\omega_{b}}{k_{B}\ln(1/\bar{n}_{b}^{\mathrm{f}}+1)}.
\end{equation}

For a simple case of $\Delta=\omega_{b}$ (namely, the driving light
is exactly resonant to the atomic transition $\left\vert
b_{0}\right\rangle \rightarrow\left\vert a_{0}\right\rangle $:
$\omega_{d}=\omega_{ab} $), the nice cooling reaches with
\begin{equation}
\bar{n}_{b}^{\mathrm{f}}=\frac{\gamma_{b}\bar{n}_{b}+\gamma_{a}\bar{n}_{a}%
}{\gamma_{a}+\gamma_{b}}\approx\frac{\gamma_{b}}{\gamma_{a}}\bar{n}_{b}%
+\bar{n}_{a}%
\end{equation}
in the strong driving strength limit $\Omega\gg\gamma_{a},\gamma_{b}$. For a
realistic atomic system, one has $\gamma_{a}\gg\gamma_{b}$ and $\bar{n}_{b}%
\gg\bar{n}_{a}$ ($\omega_{a}\gg\omega_{b}$). Especially, when $\gamma_{b}$ is
sufficiently small such that $\gamma_{b}\bar{n}_{b}\ll\gamma_{a}\bar{n}_{a}%
$~\cite{atom parameters}, the final mean population reaches its limit:
$\bar{n}_{b}^{\text{\textrm{f}}}\rightarrow\bar{n}_{b}^{\text{\textrm{lim}}%
}=\bar{n}_{a}$.

As mentioned above, the mean thermal population of mode $a$ is usually tiny,
which means that the atomic collective-excitation mode $b$ can be cooled close
to its ground state with the final thermal population $\bar{n}_{b}%
^{\text{\textrm{f}}}\rightarrow\bar{n}_{a}\ll1$.

A physical explanation of the above results can resort to the
sideband-cooling-like mechanism (see Fig.~\ref{atomic-ensemble}(b)).
The Jaynes-Cummings (JC) term ($\hat{a}^{\dag}\hat{b}$) causes the
anti-Stokes transition from $\left\vert m\right\rangle
_{a}\left\vert n\right\rangle _{b}$ to $\left\vert m+1\right\rangle
_{a}\left\vert n-1\right\rangle _{b}$, which will decay fast to the
state $\left\vert m\right\rangle _{a}\left\vert n-1\right\rangle
_{b}$. Thus, such a process makes the lower-frequency oscillator $b$
to lose one quantum and then results in its cooling. When the
anti-Stokes transition is resonantly coupled, namely,
$\Delta=\omega_{b}$, or $\Delta_{c}\equiv\omega_{ab}-\omega_{d}=0$,
the best cooling happens
with the corresponding optimal final mean population ($\bar{n}_{b}%
^{\mathrm{f}}$) given by the initial population $\bar{n}_{a}$ of
higher-frequency mode $a$. All in all, in order to reach the optimal
cooling of lower-energy collective-excitation mode $b$, the
following conditions should be satisfied: (i) strong enough pumping
light $\Omega\gg\gamma_{a},\gamma_{b}$; (ii) the resonantly driving
condition: $\Delta_{c}\equiv\omega_{ab}-\omega_{d}=0$;
(iii) $\gamma_{b}\ll\gamma_{a}$ and $\bar{n}_{a}\ll\bar{n}_{b}$ (that is,
$\omega_{b}\ll\omega_{a}$). It is notable that the above three conditions can
be met for experimentally accessible parameters of realistic atomic
systems~\cite{atom parameters}.

It is seen clearly from the above analysis that the time-dependent
coupling between two large-detuned HOs could cool down the
lower-frequency one. This cooling model is different from the
existing mechanical cooling scheme based on the optical radiation
pressure \cite{radiation-pressure cooling
experiment,Wilson-Rae:2007,Marquardt:2007,Genes:2008,Rabl2009}, with
an external laser-driving. Nevertheless, we will show below that
these two cooling schemes may be generalized to a more universal
model.

\emph{Generalized sideband cooling model of two coupled HOs.-  }A
naive cooling process could be realized when a hotter object
contacts directly with a cold one. If there exists no external
driving for two objects at the same initial temperature, it is
obviously impossible that the temperature of any one can change via
their direct interaction. But the situation changes dramatically
when we add an additional time-dependent driving or manipulate the
coupling between them to be time-dependent in largely-detuned two
coupled HOs. This kind of setup leads to a more general sideband
cooling framework.
%
\begin{figure}[th]
\includegraphics[width=6cm]{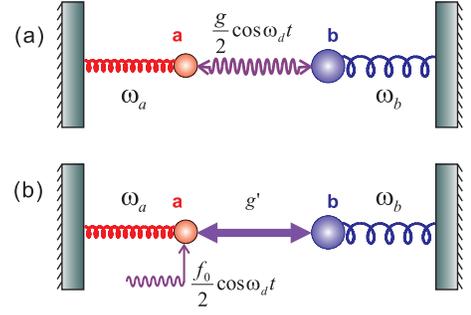}\caption{(Color online) Coupled two HOs
($a$ and $b$) model. $b$ is the desired lower-frequency HO to be cooled. (a)
The interaction between the HOs is time-dependent modulated ($\propto
g\cos(\omega_{d}t)/2$); (b) The coupling strength of the interaction between
two HOs is time-independent but there is an external time-dependent driving
($\propto f_{0}\cos(\omega_{d}t)/2$) on the higher-frequency mode $a$.}%
\label{two-HO-model}
\end{figure}

Let us first consider two coupled HOs with large-detuned frequencies
($\omega_{a}\gg\omega_{b}$) as seen in Fig.~\ref{two-HO-model}(a). The free
Hamiltonian reads $\hat{H}_{0}=\omega_{a}\hat{a}^{\dag}\hat{a}+\omega_{b}%
\hat{b}^{\dag}\hat{b}$. A time-dependent coupling is generally
expressed as
$\hat{V}_{1}(t)=g\cos(\omega_{d}t)F_{1}(\hat{a}^{\dag},\hat{a})(\hat{b}^{\dag
}+\hat{b})/2$, where $\hat{a}^{\dag}$ ($\hat{a}$) and
$\hat{b}^{\dag}$ ($\hat{b}$) are the creation (annihilation)
operators of the oscillators $a$ and $b$
with $g$ the coupling coefficient between them and $\omega_{d}$ the modulating
frequency. Here, $F_{1}(\hat{a}^{\dag},\hat{a})$ is a function of operators
$\hat{a}^{\dag}$ and $\hat{a}$. For simplicity, in what follows we consider
only the simplest case of $F_{1}(\hat{a}^{\dag},\hat{a})=\hat{a}^{\dag}%
+\hat{a}$, though a more general function (i.e., $F_{1}(\hat{a}^{\dag},\hat
{a})=\sum_{n}c_{n}\hat{a}^{\dag n}(\hat{a}^{\dag}+\hat{a})\hat{a}^{n}$ with
$c_{n}$ dimensionless coefficients) would lead to a similar result.
In the time-varying frame reference defined by
$\hat{R}(t)=\exp(-i\omega _{d}\hat{a}^{\dag}\hat{a}t)$, the
effective Hamiltonian of the coupled system reads
\begin{equation}
\hat{H}_{\mathrm{eff}}=\Delta\hat{a}^{\dag}\hat{a}+\omega_{b}\hat{b}^{\dag
}\hat{b}+g(\hat{a}^{\dag}+\hat{a})(\hat{b}^{\dag}+\hat{b}), \label{H-eff1}%
\end{equation}
where the high-oscillating terms have been neglected and the detuning
$\Delta=\omega_{a}-\omega_{d}$ could be negative when $\omega_{a}<\omega_{d}$.

Next we consider another type of two-HO system (see Fig.~\ref{two-HO-model}
(b)): a general time-independent interaction is $\hat{V}_{2}=g^{\prime}
F_{2}^{\prime}(\hat{a}^{\prime\dag},\hat{a}^{\prime})(\hat{b}^{\prime\dag
}+\hat{b}^{\prime})$ with $g^{\prime}$ the coupling strength and
$F_{2}^{\prime}(\hat{a}^{\prime\dag},\hat{a}^{\prime})$ being Hermitian, and a
periodically driving field on the higher-frequency HO reads
$\hat{H}_{d}(t)=f_{0}\cos(\omega_{d}t)(\hat{a}^{\prime\dag}+\hat{a}^{\prime
})/2$.
In the time-varying frame reference defined by $\hat{R}^{\prime}
(t)=\exp(-i\omega_{d}\hat{a}^{\prime\dag}\hat{a}^{\prime}t)$, the total
Hamiltonian reads $\hat{H}=\Delta_{0}\hat{a}^{\prime\dag}\hat{a}^{\prime
}+\omega_{b}\hat{b}^{\prime\dag}\hat{b}^{\prime}+g^{\prime}F_{2}(\hat
{a}^{\prime\dag},\hat{a}^{\prime})(\hat{b}^{\prime\dag}+\hat{b}^{\prime})$
$+f_{0}(\hat{a}^{\prime\dag}+\hat{a}^{\prime})$ with $\Delta_{0}=\omega
_{a}-\omega_{d}$ after neglecting the high-oscillating terms, where
$F_{2}(\hat{a}^{\prime\dag},\hat{a}^{\prime})$ keeps the time-independent
terms in $F_{2}^{\prime}(\hat{a}^{\prime\dag}e^{i\omega_{d}t},\hat{a}^{\prime
}e^{-i\omega_{d}t})$. Around some quasi-classical state $\left\vert
Q\right\rangle $ such that $\left\langle Q\right\vert \hat{a}^{\prime
}\left\vert Q\right\rangle =\alpha$ and $\left\langle Q\right\vert \hat
{b}^{\prime}\left\vert Q\right\rangle =\beta$, the quantum dynamics is
determined by an effective Hamiltonian $\hat{H}_{\mathrm{eff}}=\hat
{H}_{\mathrm{eff}}(\hat{a}^{\dag},\hat{b}^{\dag},\hat{a},\hat{b})$ with the
displacement operators $\hat{a}=\hat{a}^{\prime}-\alpha$ and $\hat{b}=\hat
{b}^{\prime}-\beta$ for quantum fluctuations. Then, when the displacements
$\beta$ and $\alpha$ take the equilibrium values $\beta=-F_{2}(\alpha
,\alpha)/\omega_{b}$ and $\alpha=-\left[  f_{0}+2\beta\partial_{\alpha}%
F_{2}(\alpha,y)|_{y=\alpha}\right]  /\Delta_{0}$, the effective Hamiltonian
$\hat{H}_{\mathrm{eff}}$ has the same form as that given in Eq.~(\ref{H-eff1})
with the parameters $\Delta=\Delta_{0}+2\beta\left[  \partial^{2}
F_{2}(x,y)/\partial_{x}\partial_{y}\right]  |_{x,y=\alpha}$ and $g=g^{\prime
}\partial_{\alpha}F_{2}(\alpha,y)|_{y=\alpha}$. Therefore, these types of
coupled two-HO model should have the same cooling mechanism to cool the
lower-frequency HO mode.

We wish to point out that the optical radiation-pressure cooling of
mechanical resonator
\cite{Wilson-Rae:2007,Marquardt:2007,Genes:2008,Rabl2009}, is just a
special case of the second type with $F_{2}^{\prime}
(\hat{a}^{\prime\dag},\hat{a}^{\prime})$
$=g^{\prime}\hat{a}^{{\prime}\dag }\hat{a}^{\prime}$. A similar
linearization~\cite{Wilson-Rae:2007,Rabl2009} of the effective
Hamiltonian as given in Eq.~(\ref{H-eff1}) was also mentioned in the
optical radiation-pressure cooling of mechanical resonator. Here we
present only the cooling limit (so-called sideband cooling limit)
\cite{note} of the general coupled two-HO model:
\begin{equation}
\bar{n}_{b}^{\text{\textrm{f}}}\rightarrow\bar{n}_{b}^{\text{\textrm{lim,sid}
}}=\bar{n}_{a}+\frac{\gamma_{a}^{2}}{4\omega_{b}^{2}}\approx\frac{\gamma
_{a}^{2}}{4\omega_{b}^{2}}
\end{equation}
in the resolved sideband case $\gamma_{a}^{2}\ll\omega_{b}^{2}$ when
$\Delta$ $=\sqrt{\omega_{b}^{2}+\gamma_{a}^{2}}\approx\omega_{b}$.
Here the usual relation
$\bar{n}_{a}\ll\gamma_{a}^{2}/4\omega_{b}^{2}$ has been used.

Although the above Hamiltonian~(\ref{H-eff1}) describes only a
simple coupled two-HO system, it can capture the essence of almost
all sideband cooling schemes. We need to emphasize the necessarity
of the time-dependence of modulating coupling or external driving.
It lies in a fact that, when $\omega_{a}\gg\omega_{b}$, there still
exists the effective interaction for $\left\vert \Delta\right\vert
\sim\omega_{b}$\ (or $\omega_{a}\pm\omega _{b}\sim\omega_{d}$). It
is the effective resonance $\left\vert \Delta \right\vert
\sim\omega_{b}$ that results in the sideband transitions to cool
down (or heat up) the oscillator $b$ (see
Fig.~\ref{atomic-ensemble}(b) and (c)): the JC term
($\hat{a}^{\dag}\hat{b}$) (associated with the fast decay of mode
$a$) denotes the cooling process of lower-frequency oscillator $b$
($\left\vert n\right\rangle _{b}\rightarrow\left\vert
n-1\right\rangle _{b}$); on the contrary, the anti-JC term (that is,
the anti-rotating term) ($\hat {a}^{\dag}\hat{b}^{\dag}$) denotes
the heating process of mode $b$ ($\left\vert n\right\rangle
_{b}\rightarrow\left\vert n+1\right\rangle _{b}$). When the cooling
process dominates (e.g., when $\Delta_{c}\sim0$), the cooling of
mode $b$ happens with the optimal cooling subject to the usual
sideband cooling limit
($\bar{n}_{b}^{\text{\textrm{lim,sid}}}\approx\gamma_{a}
^{2}/4\omega_{b}^{2}$).

Comparing the cooling models described by the Hamiltonians
(\ref{H-4}) and (\ref{H-eff1}), it is clear that the anti-JC terms
($\hat{a}^{\dag}\hat {b}^{\dag}+h.c.$) are absent in the former
describing the atomic ensemble. Thus, due to absence of the heating
process induced by the anti-JC term during the resolved sideband
cooling, the optimal cooling of lower-frequency
collective-excitation happens at the exact resonant ($\Delta_{c}=0$)
of (first) anti-Stokes transitions and the corresponding cooling
limit ($\bar{n} _{b}^{\text{\textrm{lim}}}=\bar{n}_{a}$) is
certainly much less than that of the usual sideband cooling limit
($\bar{n}_{b}^{\text{\textrm{lim,sid}}
}\approx\gamma_{a}^{2}/4\omega_{b}^{2}$).

\emph{Conclusion.- } We have established a theory to cool atomic
collective excitations in an optically-driven three-level atomic
ensemble. Such a cooling protocol is quite useful and promising in
quantum information processing based on atomic collective
excitations, which breaks the usual sideband cooling limit.
Moreover, motivated by the optical radiation-pressure cooling scheme
of mechanical oscillator, we have also proposed two generalized
cooling types of the coupled two-HO model: the first one possesses a
time-dependent modulating coupling coefficient between the HOs
without the external driving; while for the second one, an
additional external time-dependent driving on the higher-frequency
HO is involved, with the coupling coefficient between the HOs being
time-independent. In fact, the second type is a generalized model of
the optical radiation-pressure cooling of mechanical resonator. For
both types, the lower-frequency HO can be cooled in the resolved
sideband cooling case with the usual sideband cooling limit.

This work was supported by the RGC of Hong Kong under Grant No.~HKU7051/06P,
and partially supported by the NFRP
of China under Grant Nos.~10874091 and 2006CB921205 and NSFC Grants through
ITP, CAS.

\end{document}